\documentclass[twocolumn]{aastex6}
\usepackage{amsmath,amssymb,bm, graphicx,xspace,multirow}

\usepackage{color}
\usepackage{graphicx}
\usepackage{subfigure}
\usepackage{booktabs}

\shorttitle{baryon loading factor in Coma}
\shortauthors{Xin-Yue Shi et al.}

\begin{document}

\title{Constraining the baryon loading factor of AGN jets: implication from the gamma-ray emission of the Coma cluster}

\author{Xin-Yue Shi$^{1,2}$, Yi Zhang$^{1,2}$, Ruo-Yu Liu$^{1,2}$ and Xiang-Yu Wang$^{1,2}$}

\affil{$^1$School of Astronomy and Space Science, Xianlin Road 163, Nanjing University, Nanjing
210023, China\\
$^2$Key laboratory of Modern Astronomy and Astrophysics (Nanjing University), Ministry of Education, Nanjing 210023, People's Republic of China\\}

\begin{abstract}
High-energy cosmic rays (CRs) can be accelerated in the relativistic jets of Active Galactic Nuclei (AGNs) powered by supermassive black holes.
The baryon loading efficiency onto relativistic CR baryons from the accreting black holes is poorly constrained by observations so far. In this paper, we suggest that the $\gamma$-ray emission of galaxy clusters can be used to study the baryon loading factor of AGN jets, since  CRs injected by AGN jets are completely confined in the galaxy clusters and
sufficiently interact with intra-cluster medium via hadronic process, producing diffuse $\gamma$-rays.
We study the propagation of CRs in the galaxy clusters and calculate the radial distribution of the gamma-rays in the galaxy cluster with different injection rates from AGNs.
By comparison with the $\gamma$-ray flux and upper limits of the Coma cluster measured by $Fermi$-LAT and VERITAS, we find the upper limit of the average baryon loading factor {(defined as the efficiency with which the gravitational energy is converted into relativistic particles)} to be $\eta_{p, \mathrm{grav}} < 0.1$.
The upper limit is much lower than that required to account for diffuse neutrino flux in the conventional blazar models.
\end{abstract}



\section{introduction}

Powerful relativistic jets generated by the accretions of the supermassive black holes (SMBHs), i.e., the active galactic nuclei (AGNs), are believed to be one of the promising acceleration sites of ultra-high energy cosmic rays (UHECRs) \citep{Biermann1988, Takahara1990, Rachen1993, Berezinsky2006, Dermer2009} and high-energy neutrinos \citep{Mannheim1992, Atoyan2001, Murase2014}.
The baryon loading factor $\eta_{p,\mathrm{rad}}$ is usually defined as the the ratio between total CR
luminosity ($L_p$) and the bolometric radiation luminosity of the jet ($L_\gamma$), i.e., $\eta_{p,\mathrm{rad}}=L_p/L_\gamma$. The baryon loading factor is very relevant to the UHECR and neutrino production processes. There have been some discussions about the baryon loading factor in AGN jets based on the lepto-hadronic model of blazars \citep[e.g.,][]{Bottcher2013, Xue2019a} as well as on the potential blazar neutrinos events measured by IceCube \citep{IceCube2018, IceCube2018b, Eichmann2018, IC190730A, Giommi20, Paliya2020, Rodrigues2021}.
The accelerated protons or nuclei in blazar jets can interact with the radiation fields, which produces high-energy pions that eventually decay into photons and neutrinos. However, due to the constraints on the emission of the electromagnetic cascades by the multi-wavelength observations, the required power of relativistic protons is significantly higher than the observed radiative luminosity and even the Eddington luminosity of the SMBH \citep[e.g.][]{Keivani2018, Ansoldi2018, Gao2019, Cerruti2019, Oikonomou2019, Rodrigues2019, Zhang2020}, unless multiple radiation zones along the jet are considered \citep{Liu2019, Xue2019b, Xue2021}

In addition, if blazars can account for the diffuse neutrinos detected by IceCube, { $\eta_{p,\mathrm{rad}} \sim 3-300$ } is required with the lower limit for the case of a flat CR spectrum \citep{Murase2014}.
However, there exists a large uncertainty about this baryon loading factor so far \citep{Kadler2016, Righi2017, Palladino2019}.
In the $p\gamma$ model of blazars, the energy of relativistic protons is mainly lost through photohadronic interactions.
As the cooling timescale is much longer than the escape timescale in the AGN jets\citep{Bottcher2013, Murase2014, Gao2017}, the photohadronic interactions are very inefficient \citep{Sikora2009, Sikora2011}, and thus CRs readily escape from the the jets and host galaxy.
It is usually difficult to estimate the baryon loading factor of AGNs using the radiation of AGNs \citep{Protheroe1992,Berezhko2008}.

Galaxy clusters are the largest gravitationally bound objects in the universe.
In the galaxy clusters, CR protons injected by AGN jets are fully confined by the turbulent magnetic fields in cosmological time (\citealt{Voelk1996, Berezinsky1997}, see \citealt{Brunetti2015} for a recent review).
Hence, the confined CR protons can sufficiently interact with intra-cluster medium (ICM) via $pp$ collision to produce $\gamma$-rays and neutrinos.
Meanwhile, as galaxy clusters are large CR reservoirs which contain CRs injected in the past, the diffuse $\gamma$-rays in the present can reflect the total injected energy of CRs in history. Thus, galaxy clusters are favorable environment for constraining the efficiency of the baryon loading factor.
The gamma-ray emission from individual galaxy cluster was searched for a long time with the Fermi Large Area Telescope (LAT; \citep{Han2012, Ackermann2016}, but so far only a $\gamma$-ray source towards the Coma cluster (Abell 1656) has been detected by Fermi-LAT \citep{Xi2018, Adam2021,Baghmanyan2021}.

The Coma cluster, located at a distance of $\sim$100 Mpc (z = 0.023), is one of the nearest and most massive galaxy clusters\citep{Smith1998, Kubo2007}.
While the mass-to-energy conversion efficiency, denoted by $\epsilon$, under the standard, radiatively efficient accretion disk model \citep{Shakura1976} is $\approx 0.1$,
a higher efficiency $\epsilon \approx 0.3$ is also possible for thin-disk accretion on to a extreme Kerr black hole\citep{Thorne1974}.
Therefore, we adopt a intermediate value of $W_\mathrm{g} \approx 0.2 M_\mathrm{BH,tot} c^2 \approx 2.1 \times 10^{64} \ \mathrm{erg}$ to estimate the total gravitational energy extracted from black holes in the Coma cluster, where $M_\mathrm{BH} = 5.8 \times 10^{10} M_{\odot}$ is the total black hole mass\citep{Ensslin1998}.
As jets are powered by accretions of supermassive black holes, the total energy of AGN jets should be less than or of the same order of magnitude as the releasing gravitational potential energy.
Using the amount of the total gravitation energy, we define a new baryon loading factor ($\eta_{p, \mathrm{grav}}$) as the efficiency with which the gravitational energy is converted into relativistic particles in the present paper.

Multi-wavelength observations and analysis ranging from low frequency radio wavelengths to $\gamma$-rays centered in the direction of the Coma cluster have been reported\citep{Kent1982, Ajello2009, Brunetti2012, Xi2018, Bonafede2021}.
The Coma cluster hosts a giant radio halo\citep{Giovannini1993}, and extended soft thermal X-ray emission is observed by the ROSAT all-sky survey\citep{Briel1992}.
Using 9 years of $Fermi$ LAT data, \cite{Xi2018} reported the discovery of $\gamma$-ray emission from the Coma cluster with an extended spatial structure. The integral energy flux of $\gamma$-ray emission in the energy range of $0.2 - 300$ GeV is $\sim 2 \times 10^{-12}\ \mathrm{erg\ cm^{-2}\ s^{-1}}$ with a relatively soft spectral index of $\approx$ -2.7\citep{Xi2018}. The detection is later confirmed by the $Fermi$-LAT Collaboration\citep{Abdollahi2020, Ballet2020}, who found
a source  in the direction of the Coma cluster (named as 4FGL J1256.9+2736 in the 4FGL-DR2 catalog), as well as by other groups \citep{Adam2021,Baghmanyan2021}.
Observations with VERITAS $\gamma$-ray detector, High Energy Stereoscopic System telescopes (H.E.S.S.) and other imaging atmospheric Cherenkov telescopes have also provided upper limits on the very-high-energy (VHE) $\gamma$-ray flux of Coma cluster \citep{Perkins2006, Aharonian2009, Arlen2012, Aleksic2012}. We will use these observations to constrain the baryon loading factor of AGN jets in the Coma cluster.

This paper is organized as follows.
In Section \ref{sec:crprop}, we study the propagation of CRs in the Coma cluster and calculate the radial distribution, taking into account effects of two different injection histories.
The observed $\gamma$-ray profile and the total flux of the Coma cluster are calculated in Section \ref{sec:gamma}.
The baryon loading factors are given in section \ref{sec:baryon}.
We discuss the results and draw some conclusions in Section \ref{sec:con}.

\section{The radial density profile of CRs in the Coma cluster}
\label{sec:crprop}

Galaxy clusters are able to confine CRs for cosmological timescale \citep{Voelk1996, Berezinsky1997}.
The virial radius of the Coma cluster is $r_\mathrm{vir} \sim 3$ Mpc \citep{Lokas2003,Kubo2007}.
As the typical scale of magnetic field fluctuations $l_\mathrm{c}$ is about $1-10$\,\% of the virial radius \citep{Brunetti2015}, we have $l_\mathrm{c}\sim$ 0.3 Mpc.
The magnetic field strength of the Coma cluster can be derived from the Faraday rotation measures, $B \sim 5 \mu \mathrm{G}$ \citep{Feretti1995,Bonafede2010}.
If the Larmor radius $r_\mathrm{L}$ of a proton or nucleus with change $Z$ is smaller than the coherence length ($r_\mathrm{L} < l_\mathrm{c}$), which is satisfied with energy
$
E_\mathrm{c} \lesssim 10^{21}\ Z (B/5 \ \mathrm{\mu G})(l_\mathrm{c}/0.3\ \mathrm{Mpc})\ \mathrm{eV},
$
the propagation of CRs in the cluster turbulent magnetic field is in the diffusive regime \citep{Kotera2008}.


In the diffusive regime, the diffusion coefficient is defined as
\begin{eqnarray}
D_\mathrm{cl}
&=& \frac{1}{3} \left( \frac{B}{\delta B} \right)^2 c r_\mathrm{L}^{2-w} l_\mathrm{c}^{w-1} \nonumber\\
&\approx& 8.3 \times 10^{31}
\left( \frac{l_\mathrm{c}}{0.3 \ \mathrm{Mpc}} \right)^{2/3}
\left( \frac{E_p}{1\ \mathrm{PeV}} \right)^{1/3} \nonumber\\
&&\times \left( \frac{B Z}{5 \ \mathrm{\mu G}} \right)^{-1/3}
\ \mathrm{cm^2 s^{-1}},
\end{eqnarray}
where we have assumed
$B \sim \delta B$,
and $w=5/3$ is the spectral index for Kolmogorov diffusion.
The diffusion timescale of the CR in the Coma cluster is
\begin{eqnarray}
t_\mathrm{diff} &\approx&
\frac{r_\mathrm{vir}^2}{2D_\mathrm{cl}} \approx 16.4
\left( \frac{r_{\mathrm{vir}}}{3\ \mathrm{Mpc}} \right)^2
\left( \frac{l_{\mathrm{c}}}{0.3\ \mathrm{Mpc}} \right)^{-2/3} \nonumber\\
&&\times \left( \frac{E_p}{1\ \mathrm{PeV}} \right)^{-1/3}
\left( \frac{BZ}{5\ \mathrm{\mu G}} \right)^{1/3} \ \mathrm{Gyr}.
\end{eqnarray}
For CRs with $E_p \lesssim 0.4 \ \mathrm{PeV}$ the diffusion timescale is longer than the Hubble time $t_\mathrm{H} \sim 14\ \mathrm{Gyr}$, so these CR protons are confined in the cluster by magnetic fields.

As CRs can not escape the galaxy cluster,
the radial distribution of accumulated CRs in the Coma cluster depends on the injection history of CRs,  which determines the $\gamma$-ray production rate at present.
We consider two reference cases of CR injection history as follows:
\begin{itemize}
    \item[I]. AGN redshift evolution \\
The AGN activity changes significantly as a function of redshift, which increases with redshift up to $z \approx 2$ and then  flattens\citep{Hasinger2005, Ajello2012}.
As the CR injection is expected from the whole population of AGNs in the Coma cluster, the redshift evolution of AGN emissivities needs to be taken into consideration.
We generally assume that the injection rate in the Coma cluster follows the AGN luminosity density injected in the universe as function of redshift (or look-back time).
The luminosity functions of both X-ray-selected \citep{Hasinger2005} and Fermi-selected \citep{Ajello2012} FSRQ samples have the similar profile which peaks around redshift 1-2,
the latter shows a steeper decline after the redshift peak.
In this reference case, we use the luminosity density of Fermi-selected sample to account for the injection rate,
$Q_p(t) \propto L(t_\mathrm{lb})$,
where $t_\mathrm{lb}$ is the look-back time.

\item[II.] Constant injection rate\\
In this case, we consider a simple constant injection rate for a comparison, $Q_p(t) = const.$.
The injection time is assumed to last for $\sim 10$ Gyr.

\end{itemize}

The injected CRs at the sources are assumed to have a power-law spectrum,
$Q_p(E_p, t) \propto E_p^{-\alpha}$.
For the moment, we assume the spectrum following the one expected in the strong shocks, i.e., $\alpha \simeq 2$.

Neglecting the energy loss of particles and assuming that the particle injection occurs at $r=0$, the probability that the CRs reaches radius $r$ after time $t$ can be given by \citep{Aharonian2004}
\begin{equation}
p(E_p, r, t)= \frac{e^{-r^2/\left(4 D_\mathrm{cl}(E_p)t \right)}}{8 \pi^{3/2}\left( D_\mathrm{cl}(E_p)t \right)^{3/2}}.
\end{equation}

Therefore, the radial density distribution of the CRs can be written as
\begin{equation}
\frac{dN_p(E_p, r)}{dE_p dV} = \int_0^{t_\mathrm{max}} Q_p(E_p, t) p(E_p, r, t) dt.
\end{equation}

\begin{figure}
  \centering
  \includegraphics[width=0.47\textwidth]{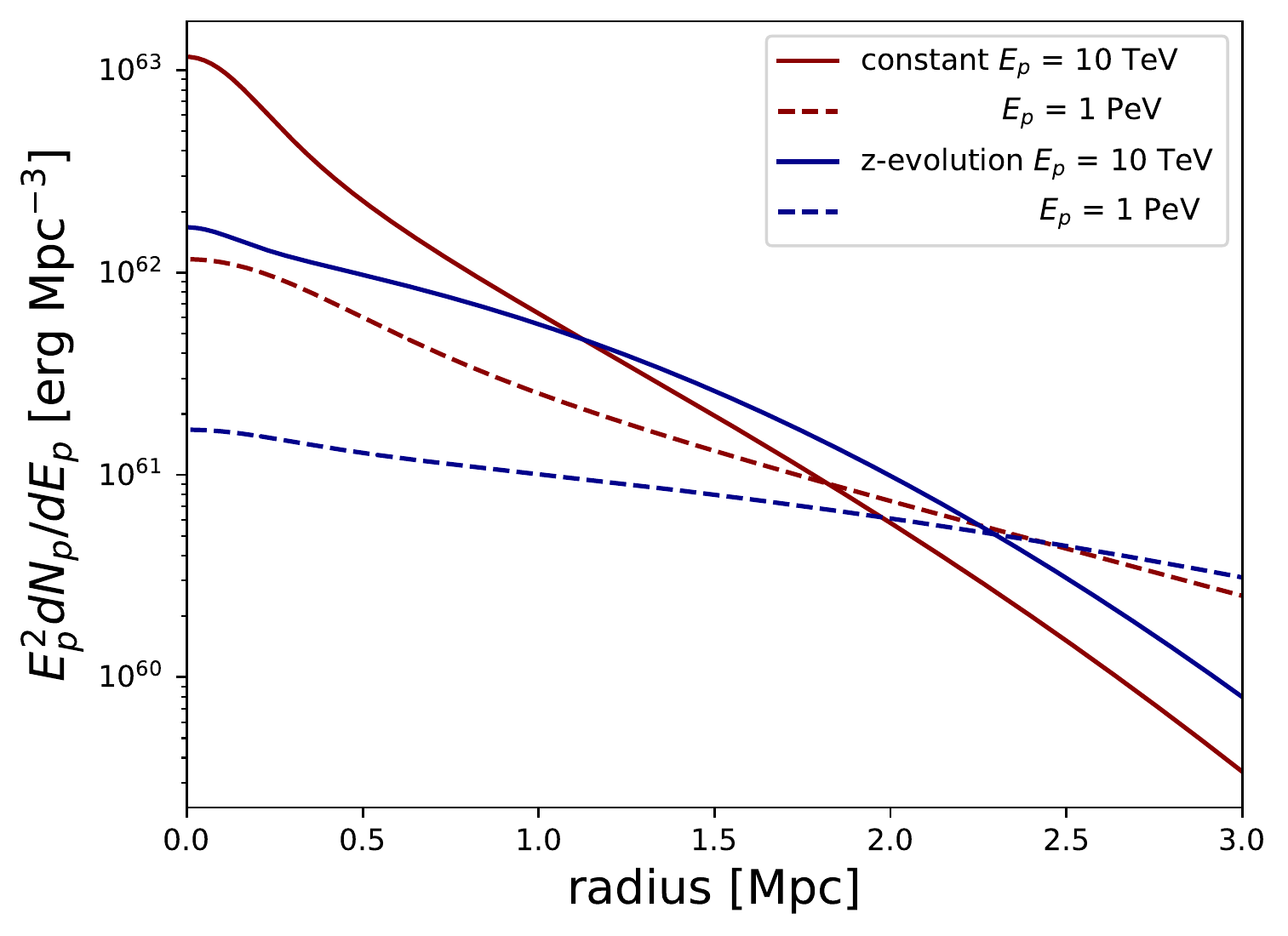}
  \caption{The density profile of CRs in the Coma cluster assuming all the gravitational energy converted into CR energy (i.e., $\eta_{p, \mathrm{grav}}=1$) and a flat CR spectrum (i.e., $\alpha=2$). The constant and redshift evolutions injections of 10 TeV (solid) and 1 PeV (dashed) CRs are shown in red and blue lines respectively.
  }
  \label{fig:pdf}
\end{figure}

\begin{figure*}[b]
  \centering
  \includegraphics[width=0.95\textwidth]{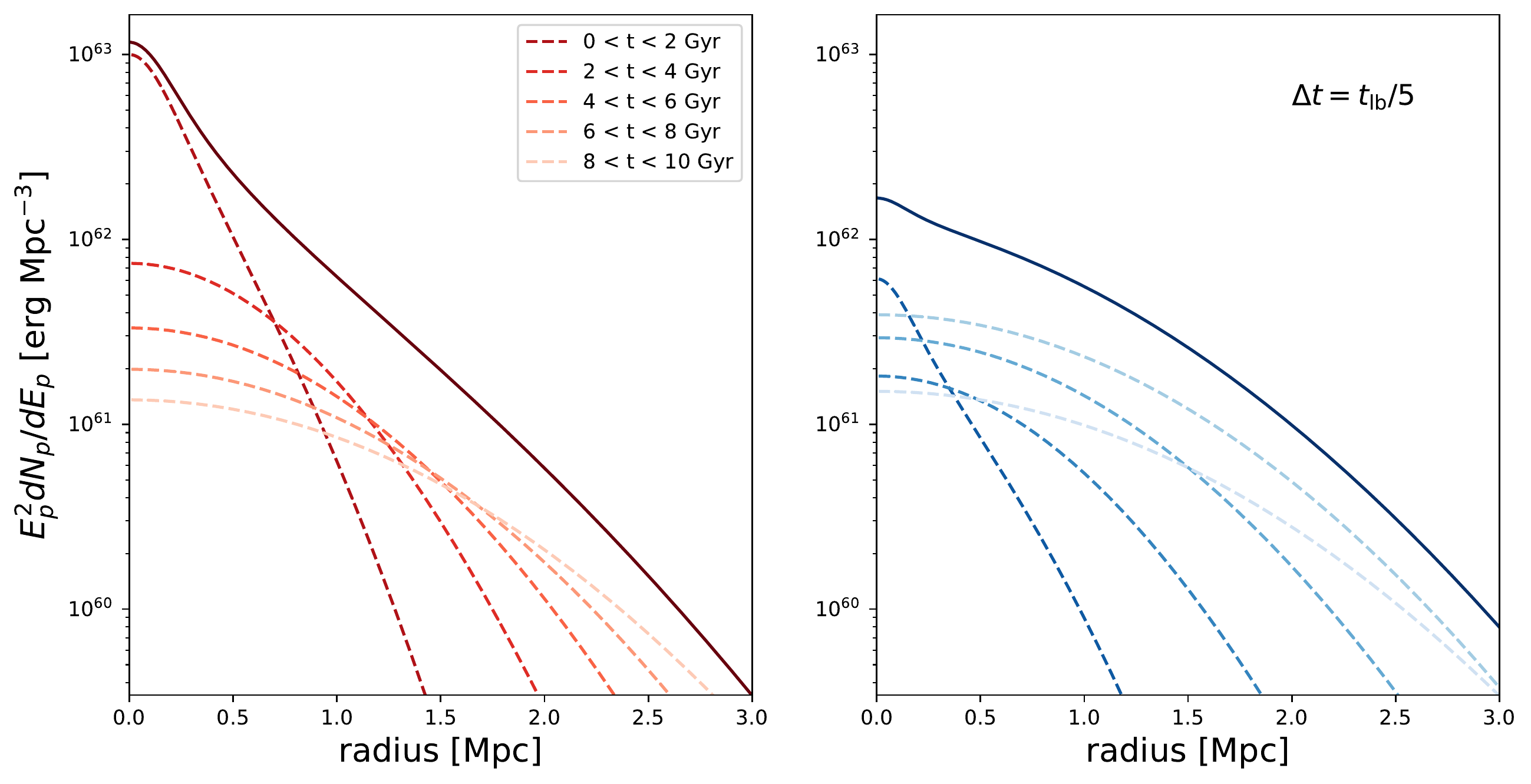}
  \caption{The solid line shows the density profile of the CRs ($E_p =10$ TeV) in the Coma cluster, which is same as Figure \ref{fig:pdf}.
  The contributions of different starting and ending times are illustrated separately by dashed lines (with different color scales).
  Left: A constant injection rate, time interval $\Delta t = 2$ Gyr. Right: The injection rate followed the redshift evolution of the AGN X-ray luminosity, time interval $\Delta t = t_\mathrm{lb,max}/5$.
  }
  \label{fig:pdf_inj}
\end{figure*}

Further assuming the total injection energy of protons equals to the total gravitational energy extracted from the central black holes, we can calculate the radial density distribution in the Coma cluster.
Figure \ref{fig:pdf} shows the obtained radial density distribution of the CRs.
As lower energy CRs diffuse more slowly than higher energy ones, the density distribution of 10\,TeV CRs is steeper than that of 1 PeV.
The time evolution of the injection rate also influences the density profile of the CRs.
In Figure \ref{fig:pdf_inj}, the contributions of different time intervals are illustrated separately using different color scales.

As redshift evolution reaches a peak at the epoch $z \sim 1-2$ \citep{Hasinger2005,Ajello2012},
if the injection rate in the cluster follows the redshift evolution, the contributions of earlier injected CRs, which are currently diffuse to farther radius, are larger than that in the case of the constant injection rate.
Therefore, for the case that the injection rate follows the redshift evolution case, the radial density distribution is flatter.
Different redshift evolutions of luminosity density have slight effect on the distribution compared to the constant case.

\section{$\gamma$-ray production}
\label{sec:gamma}

The confined CRs interact with the ICM in the cluster via the $pp$ interactions.
The products of $pp$ interaction are charged and neutral pions ($\pi^\pm$ and $\pi^0$).
Charged pions quickly decay into muons, then into electrons/positrons and neutrinos.
Neutral pions decay into $\gamma$-ray photons, $\pi^0 \rightarrow \gamma + \gamma$.
In this section, we focus on the production of $\gamma$-ray photons and calculate the  radial profile of the $\gamma$-ray flux of the Coma cluster.

Following the calculation in \cite{Kelner2006}, the $\gamma$-ray emissivity can be estimated as
\begin{eqnarray}
J_\gamma(E_\gamma, {\bf r})
&=&
\frac{dN_\gamma}{dE_\gamma dV dt} \nonumber \\
&=&
c n_\mathrm{ICM}({\bf r})
\int_{E_\gamma}^{\infty} \sigma_{pp}(E_p) \frac{dN_p(E_p, {\bf r})}{dE_p dV} \nonumber\\
&&\times F_\gamma(\frac{E_\gamma}{E_p}, E_p) \frac{dE_p}{E_p}
\end{eqnarray}
where $\sigma_{pp} (E_p)$ is the total inelastic cross section of $pp$ interactions, $F_\gamma (E_\gamma/E_p, E_p)$ is the spectrum of the secondary $\gamma$-ray in a single collision and $n_\mathrm{ICM}({\bf r})$ is the density profile of thermal gas in the ICM.

We use the classical ``$\beta$ model'' of density profile to describe the thermal gas in the ICM\citep{Cavaliere1976}, which is
\begin{equation}
n_{\mathrm{ICM}}(r) \approx n_{\mathrm{ICM}}(0) \left[ 1+ \left(r/r_\mathrm{c} \right)^2\right]^{-3\beta/2}.
\label{beta}
\end{equation}
The parameters of ICM density profile can be measured from its bremsstrahlung emission in the X-ray waveband.
For the Coma cluster, the parameters in Eq.\ref{beta} are
\begin{eqnarray}
&&n_{\mathrm{ICM}}(0) = 3.42 \times 10^{-3} \ \mathrm{cm^{-3}}, \nonumber\\
&&r_\mathrm{c} = 290\ \mathrm{kpc} \sim 0.1 r_\mathrm{vir},
\beta= 0.75,
\end{eqnarray}
which are obtained from the X-ray all-sky survey of $Roentgen$ $Satellite$ ($ROSAT$) position sensitive proportional counter (PSPC)\citep{Briel1992}.

\begin{figure}[h]
  \centering
  \includegraphics[width=0.47\textwidth]{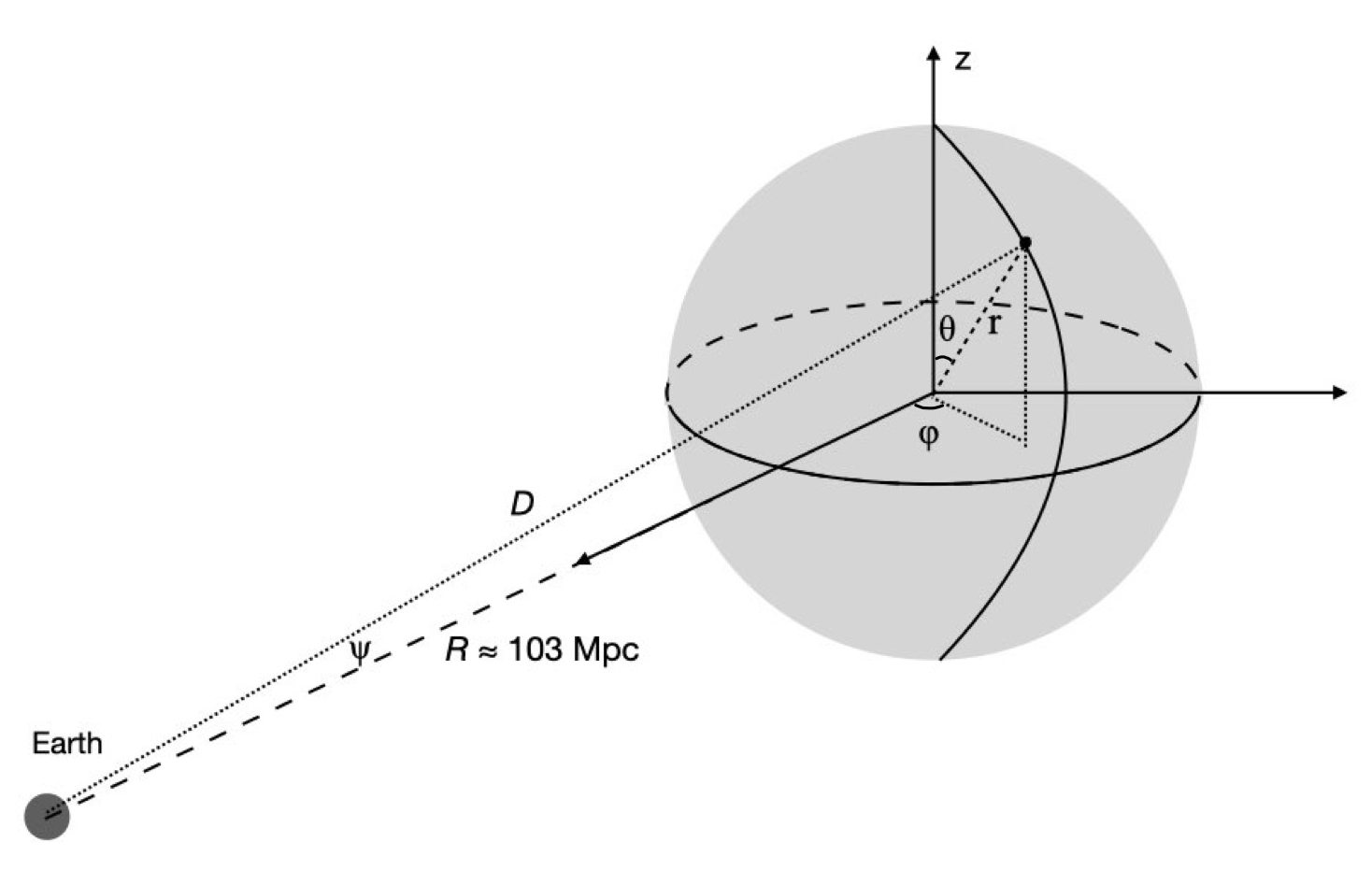}
  \caption{The diagrammatic sketch of the geometry between the Coma Cluster and Earth.
  The luminosity distance between the Coma Cluster and Earth is $R \approx 103$ Mpc.
  The distance between the element volume $dV$ with the coordinate $(r, \theta, \varphi)$ and Earth is $D$, and the viewing angle is $\psi$.}
  \label{fig:coma}
\end{figure}

Figure \ref{fig:coma} shows the geometry between the Coma cluster and Earth with a spherical coordinate system and the original point locates at the center of the Coma cluster.
The luminosity distance of Coma cluster is $R \approx 103$ Mpc, so the distance $D$ between the element volume $dV$ with coordinate $(r, \theta, \varphi)$ and the Earth is $D = \sqrt{R^2+r^2-2Rr \mathrm{sin} \theta \mathrm{cos}\varphi}$, and the viewing angle is $\mathrm{sin} \psi \approx r \mathrm{sin} \theta \mathrm{cos}\varphi/D$.
Assuming that the CR diffusion and the ICM density is the spherical symmetry distribution, the $\gamma$-ray intensity profile average over the solid angle can be calculated as
\begin{equation}
\Phi(E_\gamma, \psi) =
\frac{dN_\gamma}{dE_\gamma d\Omega dt} =
\frac{1}{2 \pi (1- \mathrm{cos} \psi_\mathrm{vir})}\int \frac{J_\gamma(E_\gamma, {\bf r})}{4 \pi {D}^2} dV,
\end{equation}
where $dV = r^2 \mathrm{sin} \theta dr d\theta d\varphi$, $2 \pi(1- \mathrm{cos} \psi_\mathrm{vir})$ is the solid angle of the $\gamma$-ray flux, $\psi_\mathrm{vir}$ is the angular distance of the virial radius.

\begin{figure}[h]
  \centering
  \includegraphics[width=0.47\textwidth]{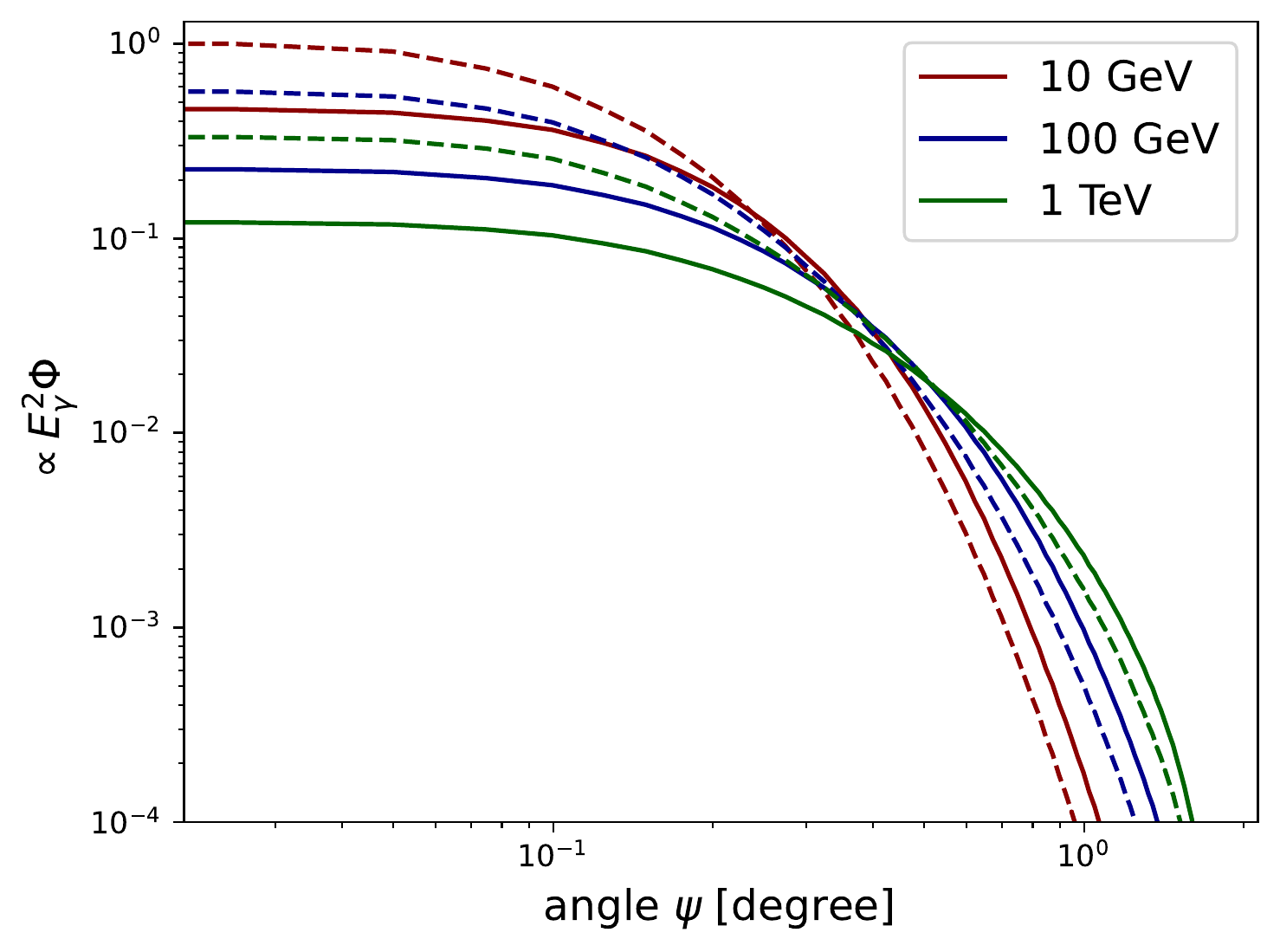}
  \caption{The normalized radial profile of $\gamma$-ray flux in terms of viewing angle $\psi$ at 10 GeV (red), 100 GeV (blue) and 1 TeV (green). The flux is normalized by fixing the flux to be unity at 10 GeV at the center of Coma cluster.
  The results of constant (dashed line) and redshift evolution (solid line) injection rates are illustrated separately.}
  \label{fig:morpho}
\end{figure}

We illustrate the gamma-ray morphologies as a function of the viewing angle $\psi$ at different photon energies in Figure \ref{fig:morpho}.
As shown in Figure \ref{fig:morpho}, the predicted $\gamma$-ray intensity profile follows both the radial density distribution of CRs and the gas. In both  cases of injection history considered here, the $\gamma$-ray intensity profiles show flat cores and steep decline at large radius.

\section{baryon loading factor in the Coma cluster}
\label{sec:baryon}

\subsection{time-averaged baryon loading factor}
\label{sec:etap}

In this section, using the total releasing gravitational potential energy $W_\mathrm{g}$ as denominator, we define the time-averaged baryon loading factor of the Coma cluster as
\begin{equation}
  \eta_{p, \mathrm{grav}} \approx \frac{W_{p,\mathrm{tot}}}{W_\mathrm{g}},
\label{eq:eta_p}
\end{equation}
where $W_{p, \mathrm{tot}}$ is the total historical injected energy of the CRs and $W_\mathrm{g}$ is the total releasing gravitational potential energy from the central black hole, $W_\mathrm{g} \approx 0.2 M_\mathrm{BH} c^2$.

The total mass of central black hole of AGNs in the Coma cluster which is proportional to the total gravitational energy $W_g$ in Eq.\ref{eq:eta_p} can also linearly effect the results.
Using $M-\sigma$ relationship, the central black hole mass was suggested to be $\sim 10^{11} M_{\odot}$\citep{Kormendy2013} given $\sigma \sim 1000 \ \mathrm{km\ s^{-1}}$ in the Coma cluster.
However, no such a massive black hole has been found at the cluster center.
The central BH mass of one of the most massive supergiant elliptical galaxy NGC 4889 is measured to be
$M_\mathrm{BH} = 2.1 \times 10^{10} M_{\odot}$\citep{McConnell2011}.
The ratio of the black hole mass to that of the spheroidal component of the stellar population of the host galaxy is estimated as $\eta_\mathrm{BH} = 0.002-0.006$\citep{Kormendy1995,Wang1998}.
In this paper, we use a conservative value $\eta_\mathrm{BH} = 0.002$, corresponding to $M_\mathrm{BH} \approx 5.8 \times 10^{10} M_\odot$ \citep{Ensslin1998} to give the constraint.

The $\gamma$-ray photons produced via the $pp$ interaction will contribute to the total gamma-ray flux from the cluster, we thus can use the observed $\gamma$-ray flux to restrict the total injection energy of protons. The absorption of gamma-ray photons by extragalactic background light (EBL) photons is not important for gamma rays below 10\,TeV from a source at 100\,Mpc. On the other hand, due to the streaming instability, CRs of energy less than tens of GeV may be dissipated by self-excited Alfv{\'e}n waves (\citealt{Kulsrud1969, Skilling1971}, see Appendix \ref{appendix:streaming} for detailed calculations in the Coma cluster), so we use the highest energy bin of $\approx$ 55 GeV in spectrum energy distributions of the extended emission component of the Coma cluster for the single radio model from \cite{Xi2018} to restrict the total injection energy of CRs at 95\% confidence level.

\begin{figure}
  \centering
  \includegraphics[width=0.47\textwidth]{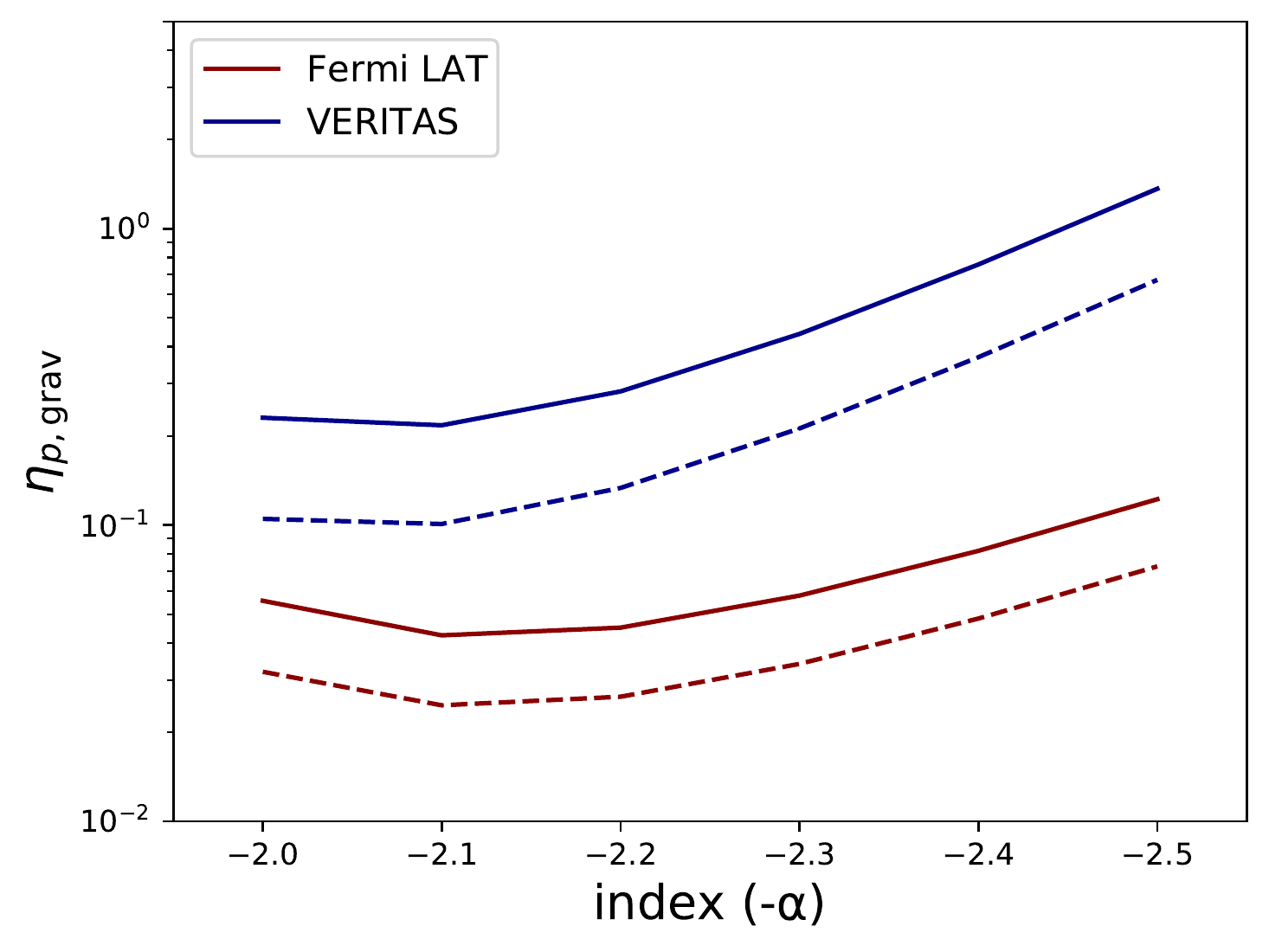}
  \caption{
  The relationship between upper limit of baryon loading factor $\eta_{p, \mathrm{grav}}$ and power-law index $\alpha$ of injected CRs for the constant (solid line) and redshift evolution (dashed line) case.
  The index range is $2 \leq \alpha \leq 2.5$.
  Results of $Fermi$-LAT (orange) and VERITAS (blue) are shown separately in different colors.
  \label{fig:etap}
  }
\end{figure}

\begin{figure}
  \centering
  \includegraphics[width=0.47\textwidth]{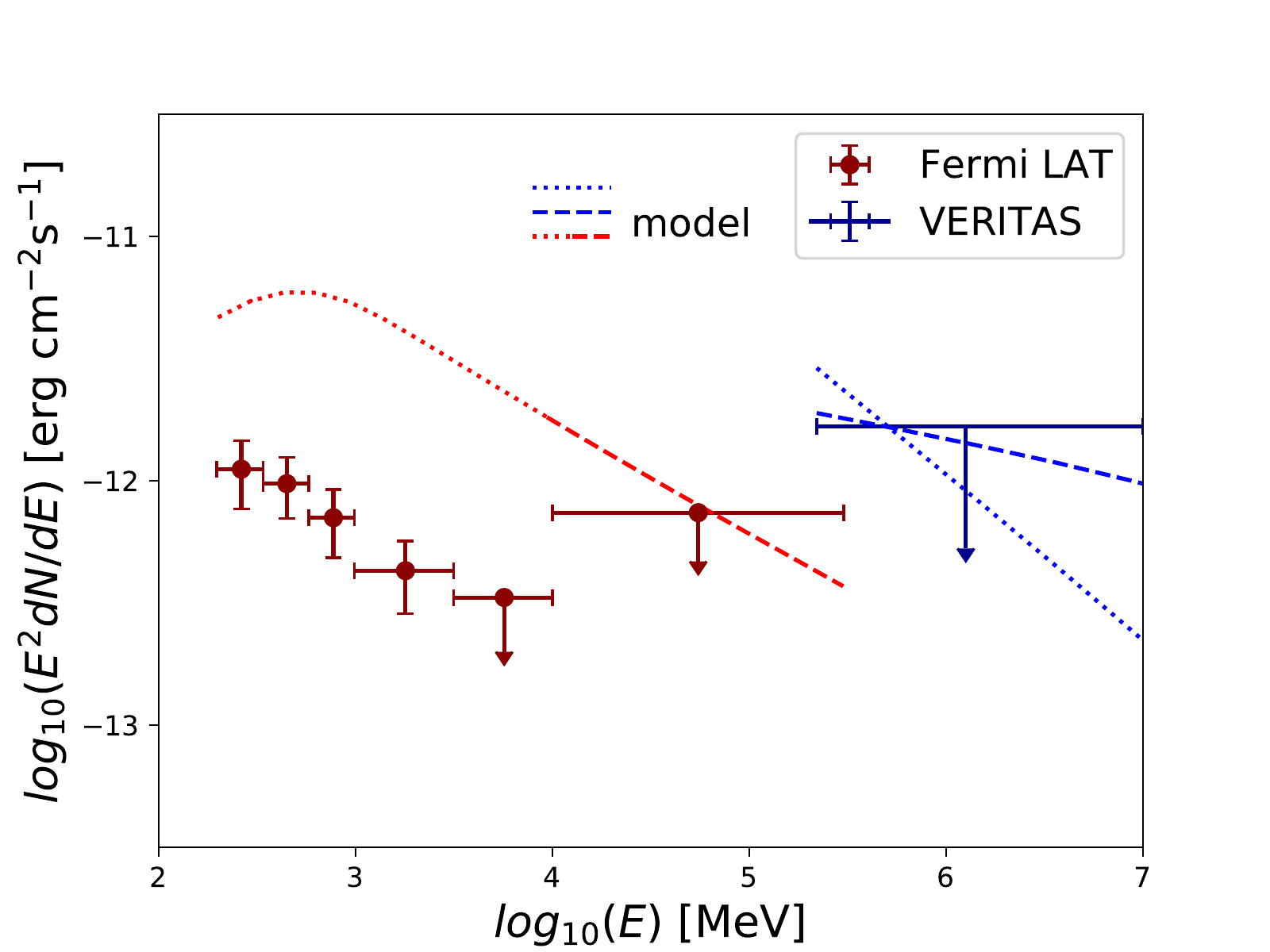}
  \caption{
  Spectrum energy distribution and upper limits for $Fermi$-LAT (red) and VERITAS (blue) observations of the Coma cluster.
  The $Fermi$-LAT data shows the extended emission component of the Coma cluster for the single radio model with integration viewing angle $\psi \approx 1.25 ^{\circ}$, corresponding to the virial radius of the Coma cluster, and the upper limits are at 95\% confidence level.
  The 99\% confidence level upper limit above an energy threshold of 220 GeV with $\alpha =2.3$ and $\psi \approx 0.4 ^{\circ}$ is shown in the VERITAS data.
  The $\gamma$-ray spectra in our model are illustrated by dashed line in the figure for comparisons.
  The best constraint using the $Fermi$-LAT data for the redshift evolution case is $\eta_{p, \mathrm{grav}} = 0.04$ and $\alpha = 2.1$ (red, energy range below the highest energy bin is illustrated by dotted line).
  Assuming that the injection spectrum index of CRs $\alpha = 2$ (dashed) and $\alpha = 2.5$ (dotted), $\gamma$-ray spectra using constraints from VERITAS observation for the redshift evolution case are also illustrated.
  \label{fig:sed}
  }
\end{figure}

Using this upper limit, we estimate the maximum allowed total injected energy of CR protons above 1 GeV and the corresponding baryon loading $\eta_{p, \mathrm{grav}}$, for two different injection histories, as well as the power-law slope of the injection spectrum $\alpha$. Since the expected pionic gamma-ray flux at a certain energy depends on both these two parameters, the obtained constraints on the two parameters are coupled.
The relationships between $\eta_{p, \mathrm{grav}}$ and $\alpha$ of protons for two reference cases are shown separately in Figure \ref{fig:etap}. The best constraint is $\eta_{p, \mathrm{grav}} \lesssim 0.02$ for the constant injection case and $ \lesssim 0.04$ for the redshift evolution case when $\alpha = 2.1$. For softer spectral indices, the upper limit increases to $\eta_{p, \mathrm{grav}} \lesssim 0.1$ for the redshift evolution case. In Figure \ref{fig:sed} we compare the observed $\gamma$-ray flux with the theoretical prediction with the parameters $\eta_{p, \mathrm{grav}} \lesssim 0.04$ (for the case of redshift evolution) and $\alpha = 2.1$. For the sake of illustration, the energy range below the the highest energy bin of $\approx$ 55 GeV is shown in dotted line and is not used for comparison, because the CR distribution at the lower energy range could be modified by the streaming instability.

Note that the streaming instability is less important at higher energy where the particle number density is lower. A more conservative constraint can be obtained by using the 99\% C.L. integral flux upper limit for $E_\gamma>220$ GeV in the Coma cluster core region ($\psi = 0.4^{\circ}$) with the 18.6 hr VERITAS observation \citep{Arlen2012}\footnote{HESS also had a measurement on the Coma cluster, obtaining a higher flux upper limit \citep{Aharonian2009}, and hence is less constraining.}. \citet{Arlen2012} provide three values of the upper limit under different assumptions of the gamma-ray index, i.e., -2.1, -2.3, and -2.5. We here take the flux upper limit with -2.3 (which is $\sim 4.7 \times 10^{-12}\ \mathrm{ph\ cm^{-2}\ s^{-1}}$) for comparison with our theoretical prediction, noting that the difference from the given upper limits for -2.1 and -2.5 is only at the level of about 10\%. The obtained constraint on $\eta_{p, \mathrm{grav}}$ with the VERITAS observation are shown in Figure \ref{fig:etap}. For softer spectral indices, the upper limit increases up to $\eta_\mathrm{p} \lesssim 1$ in the redshift evolution case. We also show the maximum $\gamma$-ray spectra allowed by the VERITAS observation for $\alpha = 2$ and $\alpha = 2.5$ in Figure \ref{fig:sed}.

In Fig.~\ref{fig:etap}, we see that the obtained $\eta_{p, \mathrm{grav}}$ with the constant CR injection rate is systematically smaller by approximately a factor of 2 than that obtained with considering the AGN redshift evolution. This is because the amount of recently injected cosmic rays is increased in the constant injection case with respect to the redshift evolution case. The recently injected CRs have not diffused to the large radius, and hence the radial distribution of CR density is steeper in the case of constant injection. Since the gas density at smaller radius is higher, this would enhance the overall gamma-ray production efficiency and thus leads to a stricter constraint on the baryon loading factor. Note that a negative evolution of blazars is observed by Fermi-LAT and X-ray selected sample, which is mainly in those low-luminosity, high-synchrotron–peaked BL Lac objects \citep{Beckmann2003, Ajello2014}. If the negative evolution is employed in the calculation, the obtained constraint on $\eta_{p, \mathrm{grav}}$ would become even stricter than that in the constant injection case, as it would further enhance the amount of recently injected CRs. However, this class of blazars constitutes only a small fraction of total blazars and is thus not representative.

In addition to blazar jets, the obtained constraints on the baryon loading factor may also apply to AGN-driven outflows, which have been suggested as high-energy particle accelerators as well \citep[e.g.][]{Jiang10, FG12, WL15, Lamastra16, Liu18}, since the energy of those powerful outflows essentially originate from the gravitational energy of matter fall into SMBHs.

\subsection{radiation related baryon loading factor}
\label{sec:etapprime}
The baryon loading factor defined in \cite{Murase2014} is
$\eta_{p,rad} = L_p/L_\gamma$,
where $L_p$ is the total CR luminosity, $L_\gamma$ is corresponding to the bolometric radiation luminosity of the jet.
In this section, we use the total radiation energy $W_\gamma$ to obtain the baryon loading $\eta_{p,rad}$ for comparison.

Assuming that the total radiation energy is proportional to the black hole mass, we first use the total released gravitational energy to normalize the integral radiation energy, $\eta_\gamma = W_\gamma/W_\mathrm{g}$.
Therefore, the baryon loading factor becomes to $\eta_{p,\mathrm{rad}} = W_\mathrm{p,tot}/W_\gamma = \eta_{p, \mathrm{grav}} / \left \langle \eta_\gamma \right \rangle$, where $\left \langle \eta_\gamma \right \rangle$ is the average fraction of the electromagnetic energy to the gravitational energy from observed blazars.

The sample is taken from \cite{Ghisellini2014}, which is composed by 191 FSRQs and 26 BL Lac objects.
Because of the lack of measurements of the black hole mass for BL Lac objects, we only use the FSRQs from the sample to calculate $\eta_\gamma$ (see Appendix \ref{appendix:eta_gamma} for detailed calculations).

\begin{figure}
  \centering
  \includegraphics[width=0.47\textwidth]{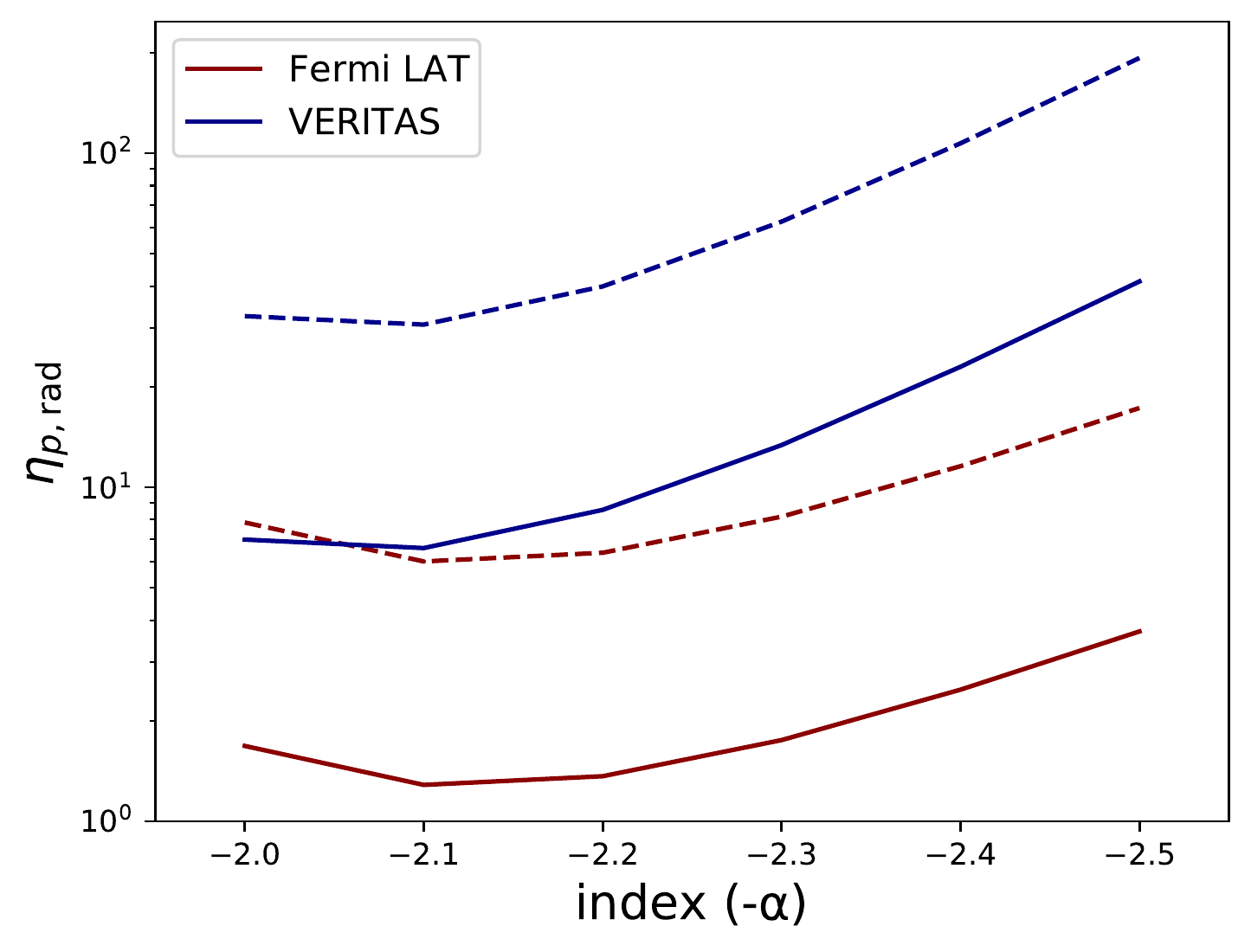}
  \caption{
  Same as Figure \ref{fig:etap}, but only $\eta_{p,\mathrm{rad}} = \eta_{p, \mathrm{grav}} / \left \langle \eta_\gamma \right \rangle$ for the redshift evolution case is shown.
  The dashed line shows results using the lower 95\% confidence range of $\left \langle \eta_\gamma \right \rangle$.  \label{fig:etap_prime}
  }
\end{figure}

The baryon loading factors with different injection indexes are shown in Figure \ref{fig:etap_prime}, $\eta_{p,\mathrm{rad}} \lesssim 1$ ($Fermi$-LAT), $\lesssim 10$ (VERITAS).
Note that in the estimation of modified baryon loading factor $\eta_{p,\mathrm{rad}}$, the same normalization with released gravitational energy of black hole mass is used in $\eta_{p, \mathrm{grav}}$ and $\eta_\gamma$.
In the calculation of $\left \langle \eta_\gamma \right \rangle$, the central black hole mass of each blazar in our sample is adopted.
However, in the calculation of $\eta_{p, \mathrm{grav}}$ in Section \ref{sec:etap}, as $W_{p,\mathrm{tot}}$ accounts for the contributions of total CR injection history in the Coma cluster, the released gravitational energy of black hole mass is estimated by the total mass of central black holes in the Coma cluster.

\section{Discussions and Conclusions}
\label{sec:con}
Galaxy clusters can effectively confine the CRs in cosmological times, so CRs can sufficiently interact with ICM to produce $\gamma$-ray and neutrino radiation.
In this paper, taking into account effects of the injection history of AGN jets, we have studied the propagation and distribution of CRs in the Coma cluster and obtained constraints on the average baryon loading factor using the $\gamma$-ray observations.
The upper limits of the average baryon loading factor are $\eta_{p, \mathrm{grav}} \sim 0.01$ and $\eta_{p, \mathrm{grav}} \sim 0.1$, respectively, from the $Fermi$-LAT and VERITAS observations for various cosmic-ray power-law indexes.
We also use the integral radiation energy to obtain the upper limits { on the conventional baryon loading factor $\eta_{p,\mathrm{rad}}$}, which are $\eta_{p,\mathrm{rad}} \sim 1$ ($Fermi$-LAT) and $\eta_{p,\mathrm{rad}} \sim 10$ (VERITAS), respectively. If such a constraint can be generalized to all the AGN in the universe, one may conclude that blazars cannot be the major sources of the diffuse neutrino background measured by IceCube, when comparing this upper limit to the theoretically required one \citep[e.g.][]{Murase2014, Palladino2019}.

Gamma-ray emission can be also produced by shocks associated with galaxy merger processes. The Coma cluster may be undergoing the merging process\citep{Tribble1993, Gurzadyan2001}. Shocks associated with galaxy merger processes would also accelerate particles to relativistic energies and produce $\gamma$-ray emission\citep{Colafrancesco1998,Ryu2003}. In this scenario, the $\gamma$-ray profile is  significantly different from the morphologies predicted by the central injection model in Figure \ref{fig:morpho} \citep{PlanckCollaboration2013, Ackermann2016}. Hence the gamma-ray profile can be used to distinguish the two scenarios for the gamma-ray emission.
Note that, as our main purpose is to give the constraint on the upper limit of baryonic loading efficiency, the $\gamma$-ray emission contributed by galaxy merger does not affect our result and can only reduce the upper limits.

In this paper, we considered the total gravitational energy of the accretion matter as the upper limit of the energy budget for accelerated protons in AGN jets or outflows. We note that an analysis of luminous blazars's spectral energy distribution based on the conventional one-zone model shows that the jet/outflow's power can be larger than the gravitational power of the accretion matter\citep{Ghisellini2014}.
The general relativistic magnetohydrodynamic (GRMHD) simulations of the radiative-inefficient accretion flow disks also shows that the jet efficiency is allowed to be greater than unity for some parameters \citep{McNamara2000}. This may not be unreasonable:
the accretion process can amplify the magnetic field, which means that the gravitational potential energy of the falling matter can be stored in the magnetic energy and be released in the certain specific state. As a result, the instant jet/outflow power may exceed the accretion power. An observational constraint for the typical AGN phase lifetime is $\sim 10^5$ yr\citep{Schawinski2015}, and the typical timescale of the blazar flare duration is from days to months.
Both timescales are much shorter compared to the total growth time $10^7 - 10^9$ yr of AGNs \citep{Fabian1999,Yu2002}, or the Salpeter time.
It implies that black holes grow via many such short bursts, so the time-averaged baryon loading factor $\eta_{p, \mathrm{grav}}$ and $\eta_{p, \mathrm{rad}}$ obtained in this paper might not apply to one short, specific status of AGNs, such as an intense flare.


Note that as the population of field galaxies is much larger than that inside clusters, the employed redshift evolution, which is obtained from the entire FSRQ population, may not be representative for AGNs in the clusters. Among many previous studies that are based on Chandra X-ray survey and the Sloan Digital Sky Survey, some suggested significant suppression of the AGN fraction in clusters than in the field \citep{Haggard2010,Hwang2012,Haines2012,Hwang2012,Ehlert2013,Ehlert2014}, but some did not\citep{Martini2013,Melnyk2013,Koulouridis2014}. Such inconsistency among different studies is probably due to the observational bias\citep{Xue2017}. If the AGN inside the cluster has a stronger redshift evolution than that in the field, the constraint on the baryon loading factor would be relaxed, while if the AGN redshift evolution is weaker in the cluster than in the field, the constraint would be stricter.

As the acceleration timescale in AGN jets is much shorter than the cooling timescales in the energy range in our study, and CRs can escape faster than they efficiently interact in jets, we use a power-law spectrum to describe the CRs leaking into the cluster. However, if CRs are still confined in the AGN jets when the acceleration shuts off, they will loss energy via adiabatic cooling due to the expansion of the jet. This would reduce the energy of cosmic rays leaking into the cluster and further increase the upper limit on the baryon loading factor.

The recent IceCube data shows that the high-energy astrophysical muon-neutrino spectrum is consistent with a single power law function \citep{Abbasi2021}.
They also test the possible existence of an additional astrophysical component following certain source-class specific flux prediction.
These tests give independent constraints on the normalization of different theoretical models, which may be converted to the constraints on the baryon loading factors. For example, the baryon loading factor $\eta_{p,\mathrm{rad}}$ of AGN inner jets model from \cite{Murase2014} is $3-300$ and is reduced to about half of the original value $\sim 1-150$ based on IceCube's constraint, which is generally in agreement with our result.

\section*{Acknowledgements}
We thank the anonymous referee for constructive comments, and Huirong Yan and Chong Ge for helpful discussions. This work is supported by NSFC grants U2031105, 11625312 and 11851304, the National Key R \& D program of China under the grant 2018YFA0404203, and China Manned Spaced Project (CMS-CSST-2021-B11).

\appendix

\section{CR streaming in the Coma cluster}
\label{appendix:streaming}

CR protons in clusters can stream along magnetic field lines down the CR gradient.
As CR stream in the plasma, a slight anisotropy in the CRs can generate unstable growth in the Alfv\'en waves, which is the so-called streaming instability\citep{Kulsrud1969, Skilling1971}.
The resulting wave growth rate can be written as \citep{Kulsrud1969}
\begin{equation}
    \Gamma_\mathrm{cr} \sim \Omega_0 \frac{n_\mathrm{cr}}{n_\mathrm{i}} \frac{v_\mathrm{D}}{v_\mathrm{A}}
\end{equation}
where $\Omega_0 = eB/(m_p c)$ is the cyclotron frequency, $n_\mathrm{cr}$ is the number density of CR protons, $n_i$ is the ion density in the plasma.
The streaming speed $v_\mathrm{D}$ can be written as $v_\mathrm{D} = D(E_p) \nabla n_{cr}/n_{cr}$, and the Alfv\'en speed is $v_\mathrm{A} = B/\sqrt{4 \pi n_\mathrm{i} m_p}$.

The wave would slow down CR diffusion and limit streaming speeds to the speed of the waves.
However, there are a variety of damping mechanisms during the scattering, including nonlinear Landau damping in the collisionless limit and wave damping by background magneto-hydrodynamic turbulence.
The non-linear Landau damping rate is \citep{Kulsrud2005}
\begin{equation}
\Gamma_\mathrm{NL} \approx 0.3 \frac{\Omega}{\mu} \frac{v_\mathrm{i}}{c} \left( \frac{\delta B}{B} \right)^2,
\end{equation}
where $\Omega = \Omega_0 /\gamma_p$, $v_\mathrm{i}$ is the velocity of thermal ions.
The turbulence damping rate is \citep{Farmer2004}
\begin{equation}
    \Gamma_\mathrm{tur} \sim \frac{v_\mathrm{A}}{\sqrt{r_\mathrm{g} l_\mathrm{A}}},
\end{equation}
where $r_\mathrm{g} = \gamma_p m_p c/(e B)$, $l_\mathrm{A} = L_\mathrm{MHD} / M_\mathrm{A}^3$, the injection scale in the ICM of Coma cluster is $L_\mathrm{MHD} = 300$ kpc, and the Alfv\'enic Mach number $M_\mathrm{A} = 10$.

\begin{figure}
    \centering
    \includegraphics[width=0.47\textwidth]{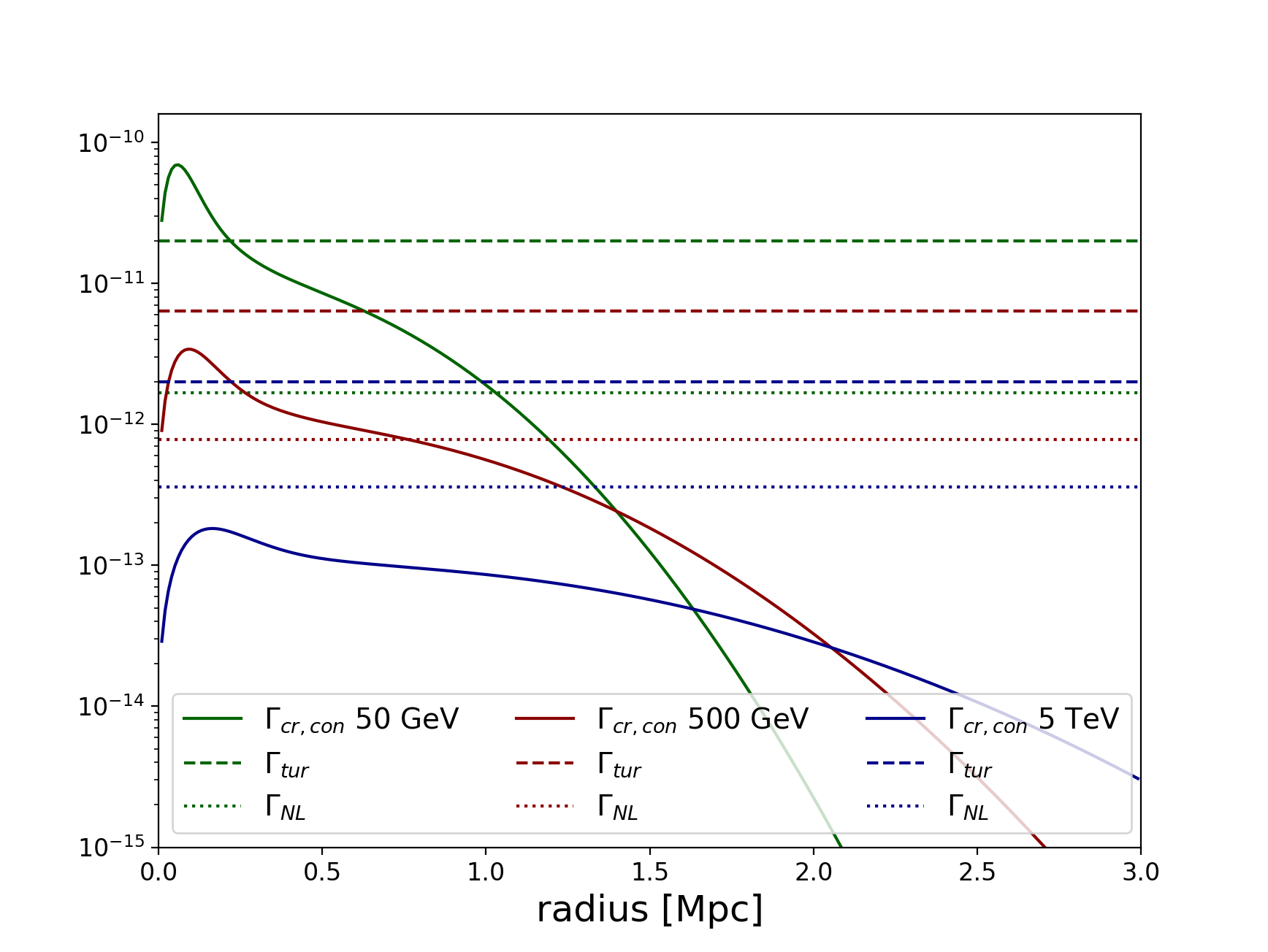}
    \caption{The growth and damping rate of CR protons in the ICM of Coma cluster. The density profile of injection protons is constant injection with power-law index $\alpha =$2 and the total injection energy is limited by Fermi observations.}
    \label{fig:gamma_cr}
\end{figure}

Figure \ref{fig:gamma_cr} shows radius distributions of the growth and damping rates  in the ICM of the Coma cluster for different proton energies.
The loading factor $\eta_{p, \mathrm{grav}}$ = 0.01 and $\alpha = 2$ for constant injection case is adopted in the figure.
When the energy of injected protons $E_p =50$ GeV, the growth of resonance wave exceeds the damping rate in the inner region of the cluster which is also the main region of $\gamma$-ray production.
When the energy of injected protons reaches 500 GeV (the corresponding energy of product gamma-ray photons is $\sim$50 GeV), the streaming instability of CR transportation in the ICM is completely suppressed by damping mechanisms, so we can use the flux upper limit of the highest-energy bin of the Fermi-LAT's data to give the constraints.

\section{calculation of the radiation energy fraction $\eta_\gamma$}
\label{appendix:eta_gamma}

In this appendix, we use a sample taken from \cite{Ghisellini2014} of 217 blazars (composed by 191 FSRQs and 26 BL Lac objects) which have been detected in the $\gamma$–ray band by the $Fermi$-LAT and have been spectroscopically observed in the optical band.
For blazars, using one–zone leptonic model, the bolometric luminosity $L_\mathrm{bol}$ can be established by multi–wavelength data\citep{Ghisellini2010,Ghisellini2014}.
The bulk Lorentz factor $\Gamma$ is determined by the viewing angle $\psi$, $\sin \psi \sim 1/\Gamma$.
Having the bolometric jet luminosity $L_\mathrm{bol}$ and the bulk Lorentz factor $\Gamma$, the absolute radiative power $P_\mathrm{rad}$ can be calculated as
$
P_\mathrm{rad} = 2 f L_\mathrm{bol} / \Gamma^2,
$
where the factor 2 accounts for two jets, $f$ is the numerical factor for external Compton process (4/3) or synchrotron and the self Compton emission (16/5).

The Salpeter time or e-folding time of an accreting black hole is $t_\mathrm{Sal} \approx 5 \times 10^7\ \mathrm{yr}$, which gives the characteristic timescale for the black hole mass to increase by one e-fold, if quasar's luminosity equals to the Eddington limit\citep{Salpeter1964}.
If the characteristic growth time of black holes is comparable to the Salpeter time, and the growth of black hole mass occurs mainly during the AGN phases,
the integral radiation energy can be estimated as $W_\gamma = P_\mathrm{rad} t_\mathrm{Sal}$.

\begin{figure}
  \centering
  \includegraphics[width=0.47\textwidth]{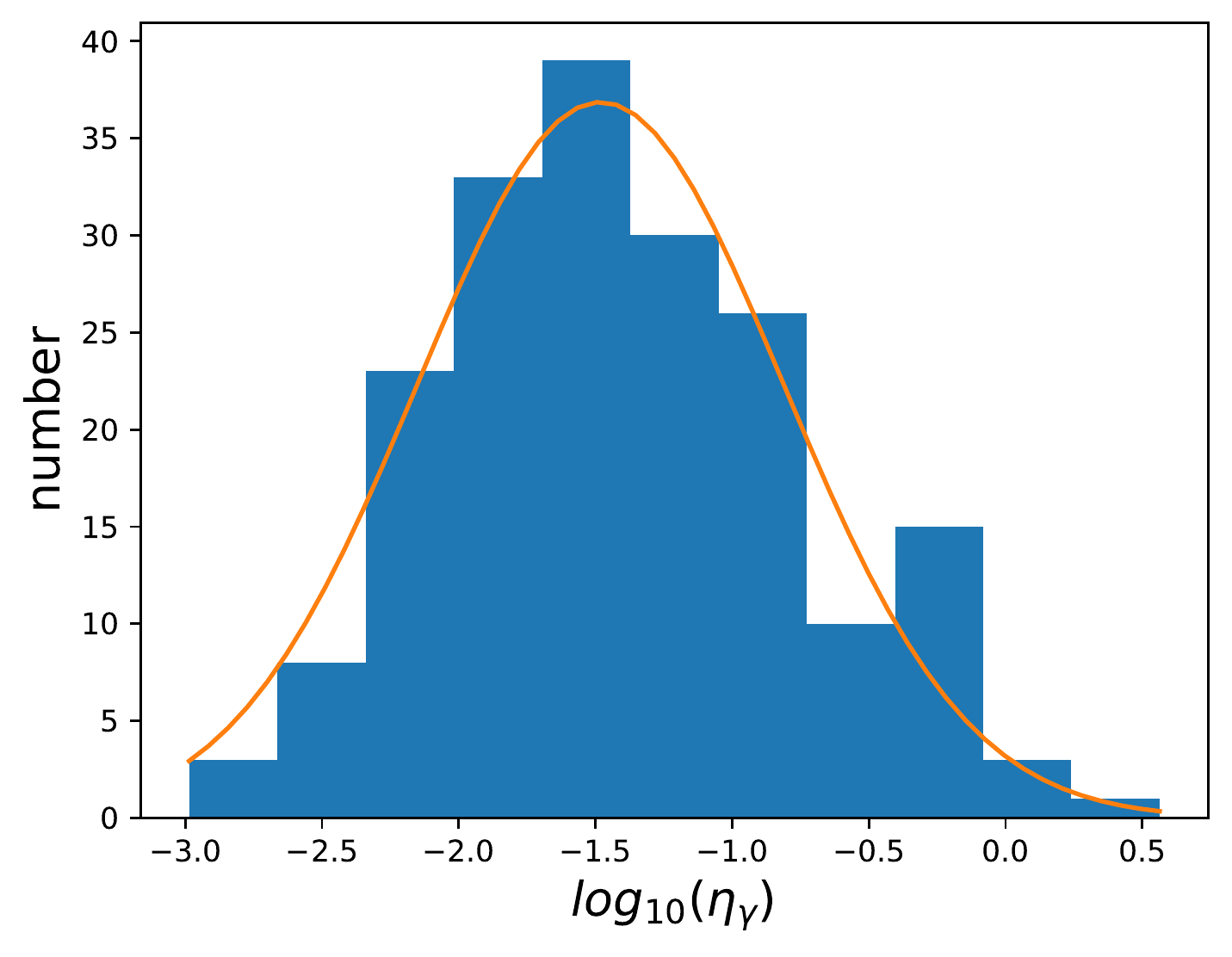}
  \caption{
  The histogram of $\eta_\gamma$ for the 191 blazars in our sample.
  The distribution (orange solid line) is fitted with a log–normal Gaussian distribution with average value $\left \langle \mathrm{log}_{10} (\eta_\gamma) \right \rangle \approx -1.48$ ($\left \langle \eta_\gamma \approx 0.03 \right \rangle$) and width $\sigma = 0.67$.\label{fig:eta_gamma}
  }
\end{figure}

The fraction of integral radiation energy to the total released gravitational energy from black holes is defined as $\eta_\gamma = W_\gamma/(0.2 M_\mathrm{BH} c^2)$,
where $M_\mathrm{BH}$ is the central black hole mass.
For FSRQs, the central black hole mass can be estimated through the virial H$\beta$, MgII and CIV broad emission lines by reverberation mapping\citep{Ghisellini2014}.
The histogram of $\eta_\gamma$ from 191 FSRQs in the sample is shown in Figure \ref{fig:eta_gamma}.
The distribution is fitted with a log-normal distribution.
The average value $\left \langle \eta_\mathrm{\gamma, FSRQ} \right \rangle \approx 0.03$, with a 95\% confidence interval $\eta_\mathrm{\gamma, FSRQ} \in [0.007, 1.5]$.
As the BL Lac objects have no central black hole mass estimation because of the lack of high S/N broad-line measurements, we use the average black hole mass of the FSRQs to give a simple estimation of 26 BL Lac objects in the sample, $\mathrm{log}_{10} (M_\mathrm{BH,avg}/M_\odot) \approx 8.5$, then $\eta_\mathrm{\gamma, BLL} \approx 0.03$, which is in good agreement with that obtained with the FSRQs.

\end{document}